\documentclass[prd,showpacs,showkeys,nofootinbib,twocolumn]{revtex4-1}

\usepackage{amsmath,amsfonts,amssymb}

\begin{document}

\def\ut{{\underset {\widetilde{\ \ }}u}}
\def\at{{\underset {\widetilde{\ \ }}a}}
\def\ap{\check{a}}
\def\ein#1{\hat{#1}}
\def\einw#1{\widehat{#1}}

\title{Equations of motion in scalar-tensor theories of gravity: A covariant multipolar approach}

\author{Yuri N. Obukhov}
\email{obukhov@ibrae.ac.ru}
\affiliation{Theoretical Physics Laboratory, Nuclear Safety Institute, 
Russian Academy of Sciences, B.Tulskaya 52, 115191 Moscow, Russia} 

\author{Dirk Puetzfeld}
\email{dirk.puetzfeld@zarm.uni-bremen.de}
\homepage{http://puetzfeld.org}
\affiliation{ZARM, University of Bremen, Am Fallturm, 28359 Bremen, Germany} 

\date{ \today}

\begin{abstract}
We discuss the dynamics of extended test bodies for a large class of scalar-tensor theories of gravitation. A covariant multipolar Mathisson-Papapetrou-Dixon type of approach is used to derive the equations of motion in a systematic way for both Jordan and Einstein formulations of these theories. The results obtained provide the framework to experimentally test scalar-tensor theories by means of extended test bodies. 
\end{abstract}

\pacs{04.25.-g; 04.50.-h; 04.20.Fy; 04.20.Cv}
\keywords{Scalar-tensor theories; Equations of motion; Variational principles}

\maketitle

%% 04.25.-g Approximation methods; equations of motion
%% 04.50.-h Higer-dimensional gravity and other theories of gravity 
%% 04.20.Fy Canonical formalism, Lagrangians, and variational principles
%% 04.20.Cv Fundamental problems and general formalism

\section{Introduction}

Scalar-tensor theories have a long history and they belong to the most straightforward generalizations of Einstein's general relativity (GR) theory. In the so-called Brans-Dicke theory \cite{Brans:Dicke:1961,Brans:1962:1,Brans:1962:2,Dicke:1962:1,Dicke:1964} a scalar field is introduced as a variable ``gravitational coupling constant'' (which is thus more correctly called a ``gravitational coupling function''). Similar formalisms were developed earlier by Jordan \cite{Jordan:1955,Jordan:1959}, Thiry \cite{Thiry:1951} and their collaborators using the 5-dimensional Kaluza-Klein approach. The interested reader may find more details on the history and developments of scalar-tensor theories in \cite{Fujii:Maeda:2003,Brans:2005,Goenner:2012,Sotiriou:2014}.

Surprisingly little attention was paid to the equations of motion of extended test bodies in scalar-tensor theories. Some early discussions can be found in \cite{Brans:1962:2,Bergmann:1968,Wagoner:1970}, and in \cite{Damour:etal:1992} the dynamics of compact bodies was thoroughly studied in the framework of the post-Newtonian formalism. However, the complete system of generalized Mathisson-Papapetrou-Dixon \cite{Mathisson:1937,Papapetrou:1951:3,Dixon:1974,Dixon:1979,Dixon:2008} equations of motion of extended test bodies in scalar-tensor theories was never derived and analyzed. Our paper fills this gap. 

\section{Point particles in scalar-tensor theories}\label{sec:point:particles}

A specific (and unusual, as compared to GR) feature of scalar-tensor theories is the status of the metric field. As it is well known, one can redefine the metric using the conformal Weyl transformation 
\begin{equation}
g_{ij}\longrightarrow \ein{g}_{ij} = \Phi^2g_{ij},\label{ggt} 
\end{equation}
and this then leads to the different form of the field equations. In this context the nomenclature {\it Jordan frame} and {\it Einstein frame} has been established in the literature, depending on whether quantities are based on the unscaled or rescaled metric in (\ref{ggt}). An interesting question then arises how the actual experiments look in these two conformally related frames. 

To illustrate this issue, let us consider the dynamics of a point particle coupled to the gravitational field. The action has the familiar form
\begin{equation}
I = \int\,ds = \int\,\sqrt{g_{ij}(x)dx^idx^j}.\label{Ids}
\end{equation}
A standard exercise is to derive the equation of motion of a point particle evaluating the variation of the action (\ref{Ids}) with respect to the particle's coordinates $x^i(s)$. The result 
\begin{equation}
{\frac {d^2x^i}{ds^2}} + \Gamma_{mn}{}^i{\frac {dx^m}{ds}}{\frac {dx^n}{ds}} = 0\label{geo}
\end{equation}
describes a geodesic curve. When such a particle is minimally coupled to the gravitational field in GR, this is the only option. Also in scalar-tensor theories we obtain the same result in Jordan's frame of reference. However, in Einstein's frame the same point particle action reads 
\begin{equation}
I = \int \Phi^{-1}\,d\ein{s} = \int\Phi^{-1}\,\sqrt{\ein{g}_{ij}(x)dx^idx^j}.\label{Idst}
\end{equation}
It is straightforward to perform a variation of the action with respect to the particle's coordinates and to derive the corresponding Euler-Lagrange equation of motion:
\begin{equation}
{\frac {d^2x^i}{d\ein{s}^2}} + \einw{\Gamma}_{mn}{}^i{\frac {dx^m}{d\ein{s}}}{\frac {dx^n}{d\ein{s}}} = - \left(\ein{g}^{ij} - {\frac {dx^i}{d\ein{s}}}{\frac {dx^j}{d\ein{s}}}\right)\partial_j\log\Phi.\label{nongeo}
\end{equation}
As we see, the dynamics of a point particle for the conformal metric $\ein{g}_{ij}$ is no longer geodetic. The particle is driven away from the geodesic curve by a ``pressure''-like force determined by the scalar field. Note that the Christoffel connection of the conformally transformed metric reads
\begin{equation}
\einw{\Gamma}_{ij}{}^k = \Gamma_{ij}{}^k + {\frac 1 \Phi}\left(\delta_i^k\partial_j\Phi + \delta_j^k\partial_i\Phi - g_{ij}\partial^k\Phi\right).\label{GG}
\end{equation}
The non-geodetic motion (\ref{nongeo}) was interpreted in earlier studies \cite{Brans:1962:1,Brans:1962:2,Dicke:1962:1,Dicke:1964} as a manifestation of a natural variability of the particle's mass due to the non-constant gravitational coupling function. 

We thus have two representations in scalar-tensor theories. If the mathematical relation between the two metrics is given, one can view the choice of $g_{ij}$ or $\ein{g}_{ij}$ as just a matter of convenience. However, a construction of a physical reference frame is a nontrivial task both for the Solar System and in cosmology, and there are controversial claims concerning the inequivalence of Einstein's and Jordan's frame in the literature, see for example \cite{Magnano:etal:1994,Capozziello:etal:2010,Corda:2011,Kamenshchik:2014}. In any case, having in mind the experimental verification of scalar-tensor theories in either Einstein's or Jordan's frame, it is necessary to derive the equations of motion for extended test bodies in scalar-tensor models. Early attempts in this direction were made in \cite{Brans:1962:2} and \cite{Damour:etal:1992}, but the complete analysis in a multipolar framework of the Mathisson-Papapetrou-Dixon type was never performed.

\section{Conservation laws in Jordan frame}

For the sake of generality, we consider the class of scalar-tensor theories along the lines of \cite{Damour:etal:1992}. Namely, we study the action 
\begin{equation}
S = \int d^4x\,{\cal L},\qquad {\cal L} = \sqrt{-g}L,\label{SJ}
\end{equation}
with the Lagrangian
\begin{eqnarray}
L &=& {\frac 1{2\kappa}}\left(- F^2R + g^{ij}\gamma_{AB}\partial_i\varphi^A\partial_j\varphi^B - 2U\right) \nonumber \\ 
&&+ L_{\rm m}(\psi,\partial\psi,g_{ij}).\label{LJ}
\end{eqnarray}
This action is an extension of the standard Brans-Dicke theory \cite{Will:1993} to the case in which we have a multiplet of scalar fields $\varphi^A$ (capital Latin indices $A,B,C = 1,\dots, N$ label the components of the multiplet). Here $\kappa = 8\pi G/c^4$ denotes Einstein's gravitational constant and in general we have several functions of scalar fields,
\begin{equation}\label{AU}
F = F(\varphi^A),\qquad U = U(\varphi^A),\qquad \gamma_{AB} = \gamma_{AB}(\varphi^A).
\end{equation}
The Lagrangian $L_{\rm m}(\psi,\partial\psi,g_{ij})$ depends on the matter fields $\psi$ and the gravitational field. 

One can understand $\ein{\kappa} := \kappa/F^2$ as a variable {\it gravitational coupling function}, and the two terms $g^{ij}\gamma_{AB}\partial_i\varphi^A\partial_j\varphi^B - 2U$ in (\ref{LJ}) determine the dynamics of the gravitational coupling.
 
Variation with respect to the scalar fields $\varphi^A$ and the metric $g_{ij}$ yields a system of field equations in the {\it Jordan reference frame}:
\begin{eqnarray}
&&\square\varphi^A + g^{ij}\gamma^A_{BC}\partial_i\varphi^B\partial_j\varphi^C + \gamma^{AB}\left(U_{,B} + F_{,B}FR\right) = 0, \nonumber \\ 
\label{phiJ}\\
&&F^2\bigl(R_{ij} - {\frac 12}Rg_{ij}\bigr) = \nabla_i\nabla_jF^2 - g_{ij}\square F^2 + \sigma_{ij} + \kappa t_{ij}. \nonumber \\ \label{gJ}
\end{eqnarray}
The $N\times N$ matrix $\gamma_{AB}$ is assumed to be non-degenerate and we denote the inverse by $\gamma^{AB}$. The subscripts ${}_{,A}$ denote derivatives with respect to the scalar fields. In particular, $F_{,A} = \partial F/\partial\varphi^A$, $U_{,A} = \partial U/\partial\varphi^A$, and we introduced
\begin{eqnarray}\label{gam3}
\gamma^A_{BC} = {\frac 12}\gamma^{AD}\left(\gamma_{BD,C} + \gamma_{CD,B} - \gamma_{BC,D}\right).
\end{eqnarray}
The energy-momentum tensor of the scalar fields 
\begin{equation}
\sigma_{ij} = \gamma_{AB}\partial_i\varphi^A\partial_j\varphi^B - {\frac 12}g_{ij}g^{kl}\gamma_{AB}\partial_k\varphi^A\partial_l\varphi^B + g_{ij}U\label{sigma}
\end{equation}
is added to the energy-momentum of matter
\begin{equation}
t_{ij} = {\frac 2{\sqrt{-g}}}\,{\frac {\delta(\sqrt{-g}L_{\rm m})}{\delta g^{ij}}},\label{tij}
\end{equation}
and together they act as sources of the gravitational field. 

Next we derive the conservation law of the energy-momentum. It is straightforward to check, using (\ref{sigma}) and the field equations (\ref{phiJ}), that 
\begin{equation}
\nabla^i\sigma_{ij} = - F_{,A}FR\partial_j\varphi^A = -{\frac 12}R\nabla_jF^2,\label{dsig}
\end{equation}
whereas using the Ricci identity we verify that
\begin{equation}
\nabla^i\left(\nabla_i\nabla_jF^2 - g_{ij}\square F^2\right) = \square\nabla_jF^2 - \nabla_j\square F^2 = R_{ij}\nabla^iF^2.\label{dA}
\end{equation}
As a result, we find that the covariant divergence $\nabla^i$ of equation (\ref{gJ}) yields the usual conservation law for matter
\begin{equation}
\nabla^i t_{ij} = 0.\label{dt}
\end{equation}

\section{Conservation laws in Einstein frame}

The conformal transformation (\ref{ggt}) with the scale factor $\Phi = F$ brings the theory to the {\it Einstein reference frame}. As a result of this transformation, the Lagrangian density in the action (\ref{SJ}) is recast into $\einw{\cal L} = \sqrt{-{\ein{g}}}\einw{L}$ with
\begin{eqnarray}
\einw{L} &=& {\frac 1{2\kappa}}\left(- \einw{R} + \ein{g}^{ij}\ein{\gamma}_{AB}\partial_i\varphi^A\partial_j\varphi^B - 2\ein{U}\right) \nonumber \\
&&+ {\frac 1{F^4}}L_{\rm m}(\psi,\partial\psi,F^{-2}\ein{g}_{ij}).\label{LE}
\end{eqnarray}
Here the scalar curvature $\einw{R}$ is constructed from the metric $\ein{g}_{ij}$, and
\begin{equation}
\ein{\gamma}_{AB} = {\frac 1{F^2}}\left(\gamma_{AB} + 6F_{,A}F_{,B}\right),\qquad 
\ein{U} = {\frac 1{F^4}}U.\label{gamUt}
\end{equation}
We will assume that the field redefinition is non-degenerate in the sense that the determinant of the resulting matrix $\ein{\gamma}_{ab}$ does not vanish. 

In the non-degenerate case, the field equations in the Einstein frame (derived from the variation of the action with respect to $\varphi^A$ and $\ein{g}_{ij}$) read
\begin{eqnarray}
&&\ein{\square}\varphi^A + \ein{g}^{ij}\ein{\gamma}^A_{BC}\partial_i\varphi^B\partial_j\varphi^C + \ein{\gamma}^{AB}\left(\einw{U}_{,B} - \frac{\kappa \ein{t}}{F^5}F_{,B}\right) = 0,\nonumber \\ 
\label{phiE}\\
&&\einw{R}_{ij} - {\frac 12}\einw{R}\ein{g}_{ij} = \ein{\sigma}_{ij} + {\frac \kappa {F^4}}\ein{t}_{ij}.\label{gE}
\end{eqnarray}
The analogs of (\ref{gam3}), (\ref{sigma}) and (\ref{tij}) in the Einstein frame are given by:
\begin{eqnarray}\label{gam3t}
\ein{\gamma}^A_{BC} &=& {\frac 12}\ein{\gamma}^{AD}\left(\ein{\gamma}_{BC,D} + \ein{\gamma}_{CD,B} - \ein{\gamma}_{BC,D}\right),\\
\ein{\sigma}_{ij} &=& \ein{\gamma}_{AB}\partial_i\varphi^A\partial_j\varphi^B - {\frac 12}\ein{g}_{ij}\ein{g}^{kl}\ein{\gamma}_{AB}\partial_k\varphi^A\partial_l\varphi^B + \ein{g}_{ij}\einw{U},\nonumber \\
\label{sigmat}\\
\ein{t}_{ij} &=& {\frac 2{\sqrt{-{\ein{g}}}}}\,{\frac {\delta(\sqrt{-{\ein{g}}}L_{\rm m})}{\delta \ein{g}^{ij}}}.\label{ttij}
\end{eqnarray}
One can verify the following relation for the trace of the energy-momentum tensor of matter:
\begin{equation}
\ein{t} = \ein{g}^{ij}\ein{t}_{ij} = g^{ij}t_{ij} = t.\label{trace}
\end{equation}

Using the scalar field equation (\ref{phiE}), we find the divergence of the scalar energy-momentum (\ref{sigmat}):
\begin{equation}
\ein{\nabla}^i\ein{\sigma}_{ij} = {\frac {\kappa\ein{t}}{F^5}}\partial_jF.\label{dsigmat}
\end{equation}
As a result, the divergence of the gravitational Einstein field equation (\ref{gE}) yields the conservation law for the energy-momentum tensor of matter
\begin{equation}
\ein{\nabla}^i\ein{t}_{ij} = {\frac 1F}\left(4\ein{t}_{ij} - \ein{g}_{ij}\ein{t}\right)\partial^iF.\label{dtt} 
\end{equation}

\section{Equations of motion in scalar-tensor theories}

The conservation laws (\ref{dt}) and (\ref{dtt}) are the starting points for the derivation of the equations of motion for extended bodies in the Jordan and in the Einstein frames, respectively. Whereas (\ref{dt}) obviously does not predict a direct influence of the scalar field on the dynamics of test particles in the Jordan frame, the conservation law (\ref{dtt}) reveals the explicit scalar field effects in the Einstein frame. 

These qualitative conclusions are completely consistent with the study of the motion of a point particle in section \ref{sec:point:particles}, where we derived (\ref{geo}) and (\ref{nongeo}) in the Jordan and in the Einstein frames, respectively. 

Our analysis of the equations of motion should add to the ongoing discussion about the Jordan and Einstein frames, and allow for the systematic testing of scalar-tensor theories by means of extended test bodies. Previously, Magnano and Sokolowski \cite{Magnano:etal:1994} argued in favor of the physicality of the Einstein frame. For other discussions of the Jordan versus Einstein frame controversy, see \cite{Catena:etal:2007,Jarv:etal:2007,Corda:2011}.

To begin with, we recast (\ref{dtt}) into an equivalent form 
\begin{equation}
\ein{\nabla}_i\ein{t}^{ij} = - A_i\left(\einw{\Xi}^{ij} + \ein{t}^{ij}\right)\label{cons}
\end{equation}
by introducing  
\begin{equation}
A_i := \partial_i\log F^{-4},\qquad \einw{\Xi}^{ij} := - \ein{g}^{ij} \ein{t}/4.\label{AZ}
\end{equation}
The equations of motion of (extended) test bodies can now be derived by means of a multipolar approximation scheme by taking the conservation law (\ref{cons}) as a starting point. 

We use the covariant expansion technique of Synge \cite{Synge:1960}, and define moments of an extended body along a reference world-line with the help of Riemannian geodesics. The general formalism was developed earlier in \cite{Puetzfeld:Obukhov:2013:1,Puetzfeld:Obukhov:2013:2,Puetzfeld:Obukhov:2013:3} in the context of gravitational theories with general nonminimal coupling to matter, and we now straightforwardly apply it to the scalar-tensor gravity. Using the proper time $\ein{s}$ to parametrize the representative world-line $y^i(\ein{s})$ of a body, we define the multipole moments as integrals over a cross-section $\Sigma(\ein{s})$ of the body's world tube, for $n = 0,1,\dots$:
\begin{eqnarray}
p^{y_1 \cdots y_n y_0}&:=& (-1)^n  \int\limits_{\Sigma(\ein{s})} \ein{\sigma}^{y_1} \cdots \ein{\sigma}^{y_n} \times \nonumber \\  && \times \ein{g}^{y_0}{}_{x_0} \sqrt{-\ein{g}}\,\ein{t}^{x_0 x_1} d \Sigma_{x_1}, \label{p_moments_def} \\
\xi^{y_2 \cdots y_{n+1} y_0 y_1}&:=& (-1)^{n} \int\limits_{\Sigma(\ein{s})} \ein{\sigma}^{y_2} \cdots \ein{\sigma}^{y_{n+1}} \times \nonumber \\  
&& \times \ein{g}^{y_0}{}_{x_0} \ein{g}^{y_1}{}_{x_1}\sqrt{-\ein{g}}\,\ein{\Xi}^{x_0 x_1} \ein{w}^{x_2} d \Sigma_{x_2}. \label{xi_moments_def}
\end{eqnarray}
Here we mark with hats the quantities related to the metric $\ein{g}_{ij}$ in the Einstein frame. In particular, $\ein{\sigma}$ denotes Synge's \cite{Synge:1960} world-function, $\ein{\sigma}^y$ denotes its first covariant derivative, and $\ein{g}^y{}_x$ is the parallel propagator with respect to the Riemannian connection in the Einstein frame.

\paragraph{Pole-dipole order}

Making use of our general results \cite{Puetzfeld:Obukhov:2013:1,Puetzfeld:Obukhov:2013:2,Puetzfeld:Obukhov:2013:3} we then obtain the equations of motion of an extended body in the pole-dipole approximation
\begin{eqnarray}
\frac{\ein{D}}{d\ein{s}} {\cal P}^a &=& \frac{1}{2} \einw{R}^a{}_{bcd} \ein{v}^b{\cal S}^{cd} - f^a, \label{dipl_eom_1}\\ 
\frac{\ein{D}}{d\ein{s}}{\cal S}^{ab} &=& - 2 \ein{v}^{[b} {\cal P}^{a]} - f^{ab}. \label{dipl_eom_2}
\end{eqnarray}
As usual, an extended body is characterized by the 4-velocity $\ein{v}^{a}=dy^{a}/d\ein{s}$ and the spin tensor $s^{ab} = 2 p^{[ab]}$. The scalar field affects the body's dynamics via the structure of the generalized momentum and angular momentum
\begin{eqnarray}
{\cal P}^a &=& F^{-4}p^a + p^{ba}\nabla_b F^{-4},\label{Pgen}\\
{\cal S}^{ab} &=& F^{-4}s^{ab},\label{Sgen}
\end{eqnarray}
and via the additional force and torque in the generalized Mathisson-Papapetrou equations
\begin{eqnarray}
f^a &=& \xi^{ab} \ein{\nabla}_bF^{-4} + \xi^{cab}\ein{\nabla}_c\ein{\nabla}_bF^{-4},\\
f^{ab} &=& 2\xi^{[ab]c}\ein{\nabla}_cF^{-4}.
\end{eqnarray}

\paragraph{Monopole order}

In the lowest approximation, we neglect the dipole moments $p^{ab}$ and $\xi^{abc}$. Then, noticing that $\xi^{ab} = - \,{\frac 14}\ein{g}^{ab}\ein{\xi}$ with $\ein{\xi} = \int\limits_{\Sigma(\ein{s})}\sqrt{-\ein{g}}\,{\ein{t}}\,\ein{w}^{x_2}d\Sigma_{x_2}$, we find that (\ref{dipl_eom_2}) yields $p^a = \ein{m}\ein{v}^a$ and $\ein{\xi} = \ein{m}$, where $\ein{m} := p^a\ein{v}_a$. Using this, we then recast (\ref{dipl_eom_1}) into the monopole equation of motion
\begin{eqnarray}
\frac{\ein{D}{\ein{v}}^a}{d\ein{s}} = - \left(\ein{g}^{ab} - \ein{v}^a \ein{v}^b\right)\ein{\nabla}_b{\rm log}F.\label{single_eom_2}
\end{eqnarray}
We thus find that the monopolar equation of motion (\ref{single_eom_2}) of an extended body is perfectly consistent with the one in (\ref{nongeo}) derived from the point particle action. 

\section{Conclusions}

We have explicitly worked out the field equations and the equations of motion for extended test bodies for a large-class of scalar-tensor theories, in the Jordan, as well as in the Einstein frame. Our results show that test bodies in the Einstein frame experience additional forces and torques, depending on the particular version of the underlying scalar-tensor theory. The mass of a body is not constant, and its dynamics is determined by the derivative of $\log F$ along a world line. This is consistent with earlier analysis \cite{Damour:etal:1992}. Furthermore, it is interesting to note, that qualitatively the structure of the equations of motion of test bodies in the Einstein frame resembles the one found in theories with nonminimal coupling to matter \cite{Puetzfeld:Obukhov:2013:1,Puetzfeld:Obukhov:2013:2,Puetzfeld:Obukhov:2013:3}. 

Our general results can be used for the systematic testing of different ``flavors'' of scalar-tensor theories. Together with our previous results in theories with nonminimal coupling \cite{Puetzfeld:Obukhov:2013:1,Puetzfeld:Obukhov:2013:2,Puetzfeld:Obukhov:2013:3}, we have now established the basis for further systematic tests and comparisons of very large classes of gravitational theories by means of extended test bodies.

\begin{acknowledgments}
This work was supported by the Deutsche Forschungsgemeinschaft (DFG) via the grant LA-905/8-1/2 and SFB 1128/1 (D.P.).
\end{acknowledgments}
\bigskip

\appendix

\section{Conventions \& Symbols}\label{conventions_app}

Our basic conventions are as in \cite{Puetzfeld:Obukhov:2013:3}. In particular, we use the Latin alphabet to label the spacetime coordinate indices. The Ricci tensor is introduced by $R_{ij} := R_{kij}{}^k$, and the curvature scalar is $R := g^{ij}R_{ij}$. Note that our curvature conventions differ by a sign from those in \cite{Synge:1960,Poisson:etal:2011}. The signature of the spacetime metric is assumed to be $(+1,-1,-1,-1)$.

\bibliographystyle{unsrtnat}
\bibliography{steoms_bibliography}

\end{document}